\documentstyle[epsfig]{mn}

\def\ref{\par\noindent\hang}

\def\ltsima{$\; \buildrel < \over \sim \;$}
\def\simlt{\lower.5ex\hbox{\ltsima}}
\def\gtsima{$\; \buildrel > \over \sim \;$}
\def\simgt{\lower.5ex\hbox{\gtsima}}
\def\simless{\mathbin{\lower 3pt\hbox
   {$\rlap{\raise 5pt\hbox{$\char'074$}}\mathchar"7218$}}}   % < or of order
\def\simgreat{\mathbin{\lower 3pt\hbox
   {$\rlap{\raise 5pt\hbox{$\char'076$}}\mathchar"7218$}}}   % > or of order
\def\ib{$I_{814}$\/ }
\def\vb{$V_{606}$\/ }
\def\ihk{$I_{814}-HK'$\/ }

\title[The Colour Distribution of Faint Field Spheroidals]{The
Optical-Infrared Colour Distribution of a Statistically-Complete
Sample of Faint Field Spheroidal Galaxies}

\author[F. Menanteau et al.]{
    F. Menanteau $^1$,
    R. S. Ellis$^1$,
    R. G. Abraham$^{1,2}$,
    A. J. Barger$^{3}$, and
    L. L. Cowie$^3$\\
  $^1$Institute of Astronomy, University of Cambridge, Madingley Road,
  Cambridge CB3 OHA, England\\
  $^2$Royal Greenwich Observatory, Madingley Road, Cambridge, CB3 0EZ,
  England\\
  $^3$Institute for Astronomy, 2680 Woodlawn Drive, Honolulu, HI
  96822, USA\\
    }

\date{Received:\ \ \ Accepted: }

\pagerange{\pageref{firstpage}--\pageref{lastpage}}

\begin{document}

\maketitle

\label{firstpage}

\begin{abstract}

In hierarchical models, where spheroidal galaxies are primarily
produced via a continuous merging of disk galaxies, the number of
intrinsically red systems at faint limits will be substantially
lower than in ``traditional'' models where the bulk of star
formation was completed at high redshifts. In this paper we
analyse the optical--near-infrared colour distribution of a large
flux-limited sample of field spheroidal galaxies identified
morphologically from archival {\em Hubble Space Telescope} data.
The $I_{814}-HK'$ colour distribution for a sample jointly limited
at $I_{814}<$23 mag and $HK'<$19.5 mag is used to constrain their
star formation history.  We compare visual and automated methods
for selecting spheroidals from our deep HST images and, in both
cases, detect a significant deficit of intrinsically red
spheroidals relative to the predictions of high-redshift
monolithic collapse models. However the  overall space density of
spheroidals (irrespective of colour) is not substantially
different from that seen locally. Spectral synthesis modelling of
our results suggests that high redshift spheroidals are dominated
by evolved stellar populations polluted by some amount of
subsidiary star formation. Despite its effect on the
optical-infrared colour, this star formation probably makes only a
modest contribution to the overall stellar mass. We briefly
discuss the implications of our results in the context of earlier
predictions based on models where spheroidals assemble
hierarchically.

\end{abstract}


\section{INTRODUCTION}

The age distribution of elliptical galaxies is a controversial issue
central to testing hierarchical models of galaxy formation. The
traditional viewpoint (Baade 1957, Sandage 1986) interprets the low
specific angular momentum and high central densities of elliptical
galaxies with their dissipationless formation at high redshift. In
support of this viewpoint, observers have cited the small scatter in
the colour-magnitude relation for cluster spheroidals at low redshifts
(Sandage \& Visvanathan 1978, Bower et al 1992) and, more recently,
such studies have been extended via HST imaging to high redshift
clusters (Ellis et al 1997, Stanford et al 1997). Examples of
individual massive galaxies with established stellar populations can
be found at quite significant redshifts (Dunlop 1997).

In contrast, hierarchical models for the evolution of galaxies
(Kauffmann et al 1996, Baugh et al 1996) predict a late redshift
of formation for most galactic-size objects because of the need
for gas cooling after the slow merger of dark matter halos.  These
models propose that most spheroidal galaxies are produced by
subsequent mergers of these systems, the most massive examples of
which accumulate since $z\simeq$1. Although examples of apparently
old ellipticals can be found in clusters to quite high redshift,
this may not be at variance with expectations for hierarchical
cold dark matter (CDM) models since clusters represent regions of
high density where evolution might be accelerated (Governato et al
1998). By restricting evolutionary studies to high density
regions, a high mean redshift of star formation and homogeneous
rest-frame UV colours would result; such characteristics would not
be shared by the field population.

Constraints on the evolution of field spheroidals derived from
optical number counts as a function of morphology (Glazebrook et
al 1995, Im et al 1996, Abraham et al 1996a) are fairly weak,
because of uncertainties in the local luminosity function.
Nonetheless, there is growing evidence of differential evolution
when their properties are compared to their clustered
counterparts. Using a modest field sample, Schade et al (1998)
find a rest-frame scatter of $\delta(U-V)$=0.27 for distant
bulge-dominated objects in the HST imaging survey of CFRS/LDSS
galaxies, which is significantly larger than the value of
$\simeq$0.07-0.10 found in cluster spheroidals at $z\simeq$0.55 by
Ellis et al 1997. Likewise, in their study of a small sample of
galaxies of known redshift in the {\em Hubble Deep Field} (HDF),
Abraham et al (1998) found a significant fraction ($\simeq$40\%)
of distant ellipticals showed a dispersion in their internal
colours indicating they had suffered recent star formation
possibly arising from dynamical perturbations.

Less direct evidence for evolution in the field spheroidal
population has been claimed from observations which attempt to
isolate early-type systems based on predicted colours, rather than
morphology. Kauffmann et al (1995) claimed evidence for a strong
drop in the volume density of early-type galaxies via a
$V/V_{max}$ analysis of colour-selected galaxies in the {\em
Canada-France Redshift Survey} (CFRS) sample (Lilly et al 1995).
Their claim remains controversial (Totani \& Yoshii 1998, Im \&
Castertano 1998) because of the difficulty of isolating a robust
sample of field spheroidals from $V-I$ colour alone (c.f. Schade
et al 1998), and the discrepancies noted between their analyses
and those conducted by the CFRS team.

In addition to small sample sizes, a weakness in most studies of high
redshift spheroidals has been the paucity of infrared data.  As shown
by numerous authors (eg. Charlot \& Silk 1994), near-IR observations
are crucial for understanding the star formation history of distant
galaxies, because at high redshifts optical data can be severely
affected by both dust and relatively minor episodes of
star-formation. Recognizing these deficiencies, Moustakas et al (1997)
and Glazebrook et al (1998) have studied the optica-infrared colours
of small samples of of morphologically-selected galaxies. Zepf (1997)
and Barger et al (1998) discussed the extent of the red tail in the
optical-IR colour distribution of HDF galaxies. Defining this tail
($V_{606}$-$K>$7 and $I_{814}$-$K>$4) in the context of evolutionary
tracks defined by Bruzual \& Charlot's (1993) evolutionary models,
they found few sources in areas of multicolour space corresponding to
high redshift passively-evolving spheroidals.

The ultimate verification of a continued production of field
ellipticals as required in hierarchical models would be the
observation of a decrease with redshift in their comoving space
density. Such a test requires a large sample of
morphologically-selected ellipticals from which the luminosity
function can be constructed as a function of redshift. By probing
faint in a few deep fields, Zepf (1997) and Barger et al (1998) were
unable to take advantage of the source morphology; constraints derived
from these surveys relate to the entire population. Moreover, there is
little hope in the immediate term of securing spectroscopic redshifts
for such faint samples. The alternative adopted here is to combine
shallower near-infrared imaging with more extensive HST archival
imaging data, allowing us to isolate a larger sample of {\it brighter,
morphologically-selected} spheroidals where, ultimately, redshifts and
spectroscopic diagnostics will become possible. Our interim objective
here is to analyse the optical-infrared colour distribution of faint
spheroidals which we will demonstrate already provides valuable
constraints on a possible early epoch of star formation.

A plan of the paper follows. In $\S$2.1 we discuss the available HST
data and review procedures for selecting morphological spheroidals
from the images. In $\S$2.2 we discuss the corresponding ground-based
infrared imaging programme and the reduction of that data. The merging
of these data to form the final catalogue is described in $\S$2.3. In
$\S$3 we discuss the optical-infrared colour distribution for our
sample in the context of predictions based on simple star formation
histories and consider the redshift distribution of our sample for
which limited data is available. We also examine constraints based on
deeper data available within the Hubble Deep Field. In $\S$4 we
summarise our conclusions.

\section{CONSTRUCTION OF THE CATALOGUE}

\subsection{THE HST SAMPLE}

\begin{figure*}
\begin{center}
\leavevmode
\centerline{\psfig{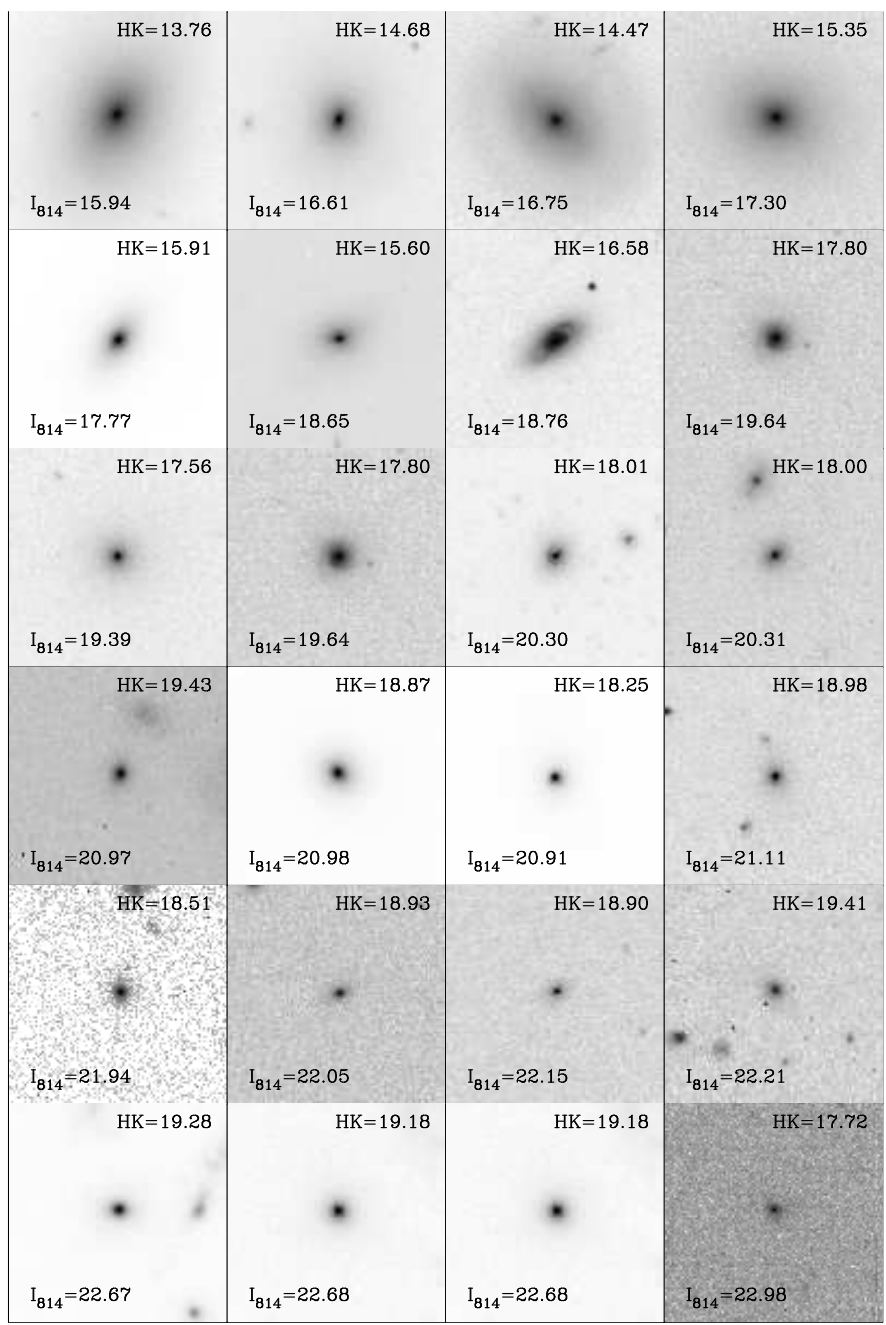}}
\end{center}
\caption{\em A selection of visually-classified spheroidal galaxies
sorted by \ib magnitude, selected from the HST archive. Each panel
represents 10 arcsec on a side.}
\label{fig:Figure1}
\end{figure*}

In searching the HST archive for suitable fields, we adopted a minimum
$I$ F814W-band exposure time of 2500 sec and a Galactic latitude of
$|b|$=19$^{\circ}$ so that stellar contamination would not be a major
concern. These criteria led to 48 fields accessible from the Mauna Kea
Observatory comprising a total area of 0.0625 deg$^2$(225 arcmin$^2$).
Table 1 lists the fields adopted, including several for which limited
redshift data is available e.g. the HDF and its flanking fields
(Williams et al 1996), the Groth strip (Groth et al 1994) and the
CFRS/LDSS survey fields (Brinchmann et al 1997). F606W imaging is
available for 25 of the fields in Table~1.  Object selection and
photometry for each field was performed using the {\tt SExtractor}
package (Bertin \& Arnouts 1996). Although the detection limit varies
from field to field, the \ib-band data is always complete to $\sim 24$
mag and the \vb-band to $\sim 25$ mag\footnote{These and subsequent
detection limits refer to near-total magnitudes in the Vega system
based on profile fitting within the SExtractor package
(`$m_{best}$').}.

The morphologies of galaxies in the sample were investigated
independently using visual classifications made by one of us
(RSE), and automated classifications based on the central
concentration ($C$) and asymmetry ($A$) parameters defined in
Abraham et al. (1996b). In the case of visual classifications we
adoped the MDS scheme defining spheroidal to include E: E/S0: S0
and S0/a (MDS types 0,1,2). As shown below, the visual and
automated classifications compare quite favourably, with
particularly satisfactory agreement for the regular spheroidal
galaxies that are the focus of this paper.

\begin{figure*}
\begin{center}
\leavevmode
\centerline{\psfig{file=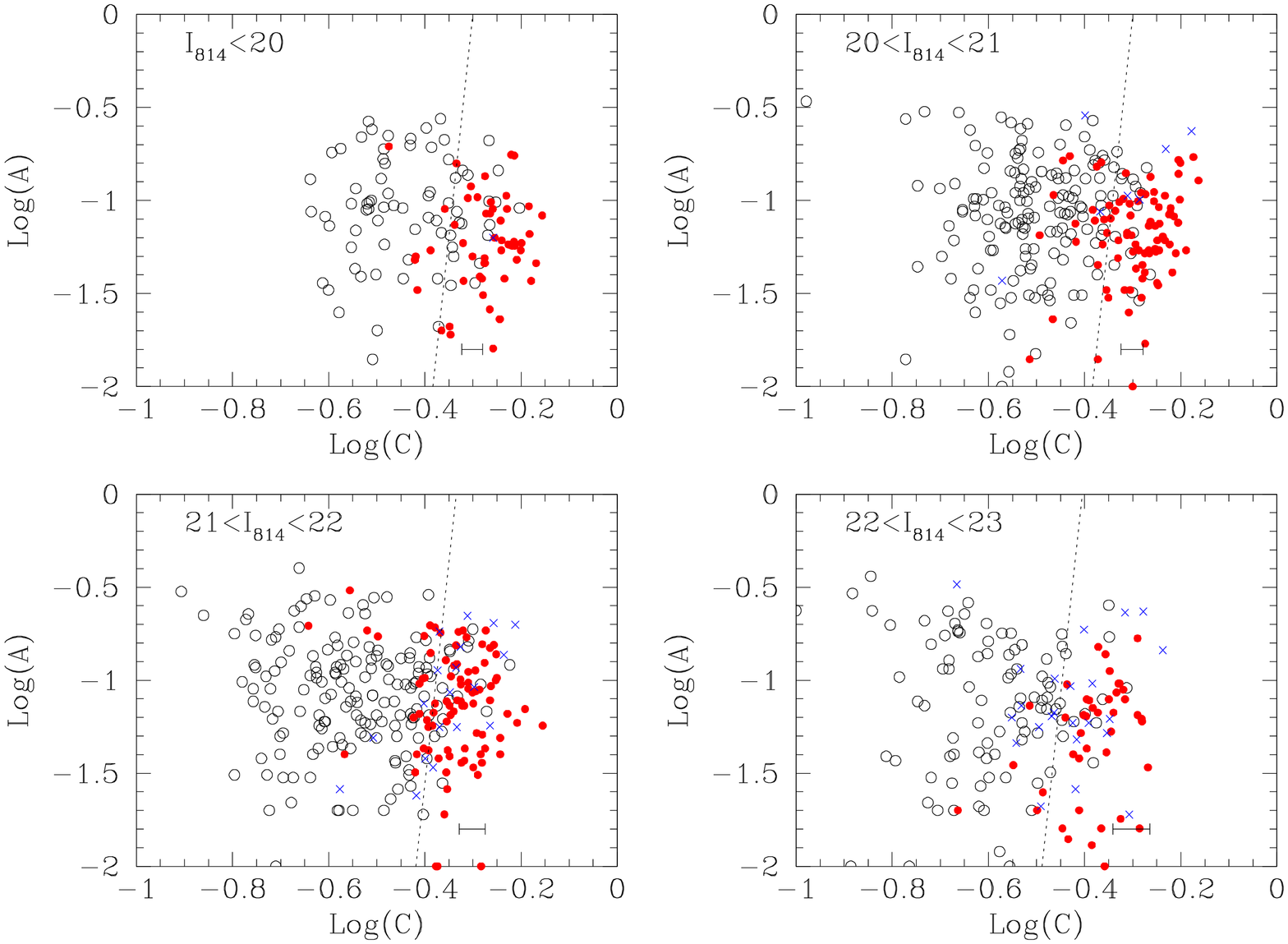,width=13cm,angle=-90}}
\end{center}
%\vspace{8cm}
\caption {\em The distribution of asymmetry and concentration for
visually-classified galaxies a) [top left] \ib$<20$ mag, b) [top right]
20$<$\ib$<$21 mag, c) [bottom left] 21$<$\ib$<$22 mag (the limit achieved
with the Medium Deep Survey) and (d) [bottom right] 22$<$\ib$<$23 mag.
The dashed lines represent optimal boundaries for the separation of
spheroidal galaxies (solid circles) from spirals and irregulars (open
circles). Compact objects are shown as crosses. The reference boundary
line is defined using the visually identified sample at \ib$<$20
and then shifted slightly as a function of magnitude on the basis of
simulations. The expected 1$\sigma$ RMS errors on measures of central
concentration are shown by the error bar on the lower right portion of
each panel.}
\label{fig:Figure2}
\end{figure*}

The appropriate limiting magnitude of our survey is set by that at
which we believe we can robustly isolate spheroidal galaxies from
compact HII galaxies, stars and bulge-dominated spirals.  The
Medium Deep Survey (MDS) analyses adopted a limiting magnitude for
morphological classification (using nine classification bins) of
\ib=22 mag, although some MDS papers extended this further to
\ib=23 mag (see Windhorst et al 1996 for a summary). In Abraham et
al (1996b) and Brinchmann et al (1997), HST data similar to that
in the present paper was also used to classify galaxies to \ib=22
mag. However, by restricting our analysis in the present paper to
spheroidal systems, we are able to extend classifications to
slightly deeper limits (\ib=23.0 mag). This is possible because
the chief diagnostic for discriminating spheroidals is central
concentration, rather than asymmetry which is sensitive to lower
surface brightness features. Because the classifications based on
$A$ and $C$ are objective, the classification limits for the
present dataset have been investigated using simulations, as
described below.

Figure~1 shows a typical set of morphologically-identified
spheroidals at various magnitudes down to \ib=23 mag. Figure~2
compares the $A-C$ and visual morphological distributions at a
range of magnitude intervals, down to the limits of our survey.
The demarcation between early and late-types on the basis of $A$
and $C$ is made using bright galaxies (\ib$<20$ mag) and shifted
slightly as a function of magnitude on the basis of simulations
made using the {\tt IRAF} package {\tt artdata}, which model the
apparent change in the central concentration of an $r^{1/4}$ law
elliptical galaxy as a function of decreasing signal-to-noise.
Random errors on central concentration are also determined on the
basis of simulations, and representative error bars are shown in
Figure~2.  The general agreement between the visual and automated
classifications is remarkably good, particularly to \ib$=$22 mag.
Between \ib$=$22 and 23 mag the agreement worsens, mostly because
of the great increase in the number of visually-classified
``compact'' systems. We define compacts to be those systems where
there is no clear distinction between small early-type galaxies,
faint stars and/or HII regions.

In order to quantify the concordance between the visual and
automated classifications, the $A-C$ distribution was analysed
using a statistical bootstrap technique (Efron \& Tibshirani
1993). The $A-C$ distribution was resampled 500 times in order to
determine the uncertainties in both the number of systems
classified as early-type, and the uncertainties on the  colour
distribution for these systems. These measurements will be
discussed further below in \S3.3.

The somewhat larger number (323 vs 266) of $A/C$-classified
early-types relative to the visually classified galaxies is
significant at the 3$\sigma$ level. However, if the compact
systems are included in the tally of visually classified
early-type systems, then the number of visually and A/C classified
ellipticals agree closely (to within 1 sigma). It is clear that
the distinction between compact galaxies and early-type systems is
an important consideration when determining the number counts of
early type systems at the faint limits of these data. However, it
is perhaps worth noting at this stage that another bootstrap
analysis (presented in \S3.3) shows that the uncertainty
introduced by compact systems into the number counts at $22<I<23$
does {\em not} manifest itself as a large uncertainty in the
colour histograms of the early-type population.

\subsection{GROUND-BASED INFRARED IMAGING}

Although some of the fields in Table 1 have \vb and \ib HST data,
such a wavelength baseline is not very useful at thesed depths. As
discussed by Moustakas et al (1997) and Zepf (1997), the addition
of infrared photometry is especially helpful in distinguishing
between passively-evolving systems and those undergoing active
star formation, {\it regardless of redshift}, primarily because of
its reduced sensitivity to K-dimming, small amounts of star
formation and dust reddening.

Our infrared imaging was mainly conducted using the QUIRC 1024$^2$
infrared imager on the University of Hawaii 2.2-m telescope.  The log
of observations is summarised in Table 2. In order to improve the
observing efficiency in securing deep infrared photometry for a large
number of WFPC-2 fields, we used the notched $H+K'$ 1.8$\mu$m filter
(which we refer to hereafter as the $HK'$ filter) (Wainscoat \& Cowie
1998, Figure~3) which offers a gain in sensitivity of typically a
factor of $\simeq$2 over a conventional $K'$ filter. At the f/10 focus
of the UH 2.2m, the field of view is $193''\times 193''$ with a scale
of $0.1886''\,$pixel$^{-1}$ ensuring that the 3 WFPC2 chips can be
comfortably contained within a single exposure.

\begin{figure}
\begin{center}
\leavevmode
\centerline{\psfig{file=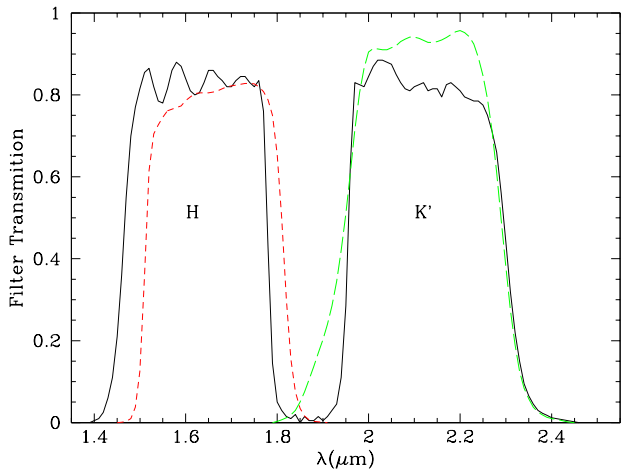,width=6.5cm,angle=270}}
\end{center}
%\vspace{6cm}
\caption{\em Transmission curve for the notched $HK'$ filter(solid
line) compared with that for the standard $H$ (dashed line) and $K'$
(long-dashed line) passbands}
\label{fig:Figure3}
\end{figure}

Each $HK'$ exposure was composed of 13 sub-exposures of $\simeq$100
sec duration (depending on the background level) spatially-shifted by
increments of 5-20 arcsec in all directions. This dithering pattern
was repeated 2-3 times during the exposure. The data was processed
using median sky images generated from the disregistered exposures and
calibrated using the UKIRT faint standards system. Most of the data
was taken under photometric conditions; deep non-photometric data was
calibrated via repeated short exposures taken in good conditions. The
limiting magnitude of the infrared data varies slightly from field to
field and is deepest for the HDF and flanking fields which were taken
in a separate campaign (Barger et al 1998).

\begin{figure}
\begin{center}
\leavevmode
\centerline{\psfig{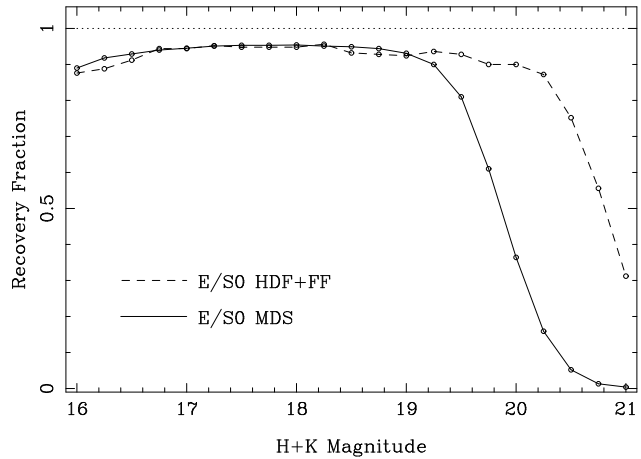}}
\end{center}
\caption {\em Statistical completeness of the UH 2.2m QUIRC data for
the HST archival fields and the HDF as determined by procedures
described in the text.}
\label{fig:figure4}
\end{figure}

In order to determine the detection limit of our $HK'$ data we
performed extensive Monte Carlo simulations. Using the IRAF {\tt
artdata} package we created simulated data sets, which were
subsequently analysed using the same extraction and measurement
methods as for the real data. With the task {\tt mkobjects} we
generated artificial galaxies assuming exponential disk profiles with
no internal absorption for spirals and de Vaucouleurs profiles for
spheroidals. The profile scales were chosen to be magnitude-dependent
converging to the image seeing at faint limits. Figure~4 shows the
results of this exercise. The 80\% completeness limit for spheroidals
is $HK'$=19.5 mag for most of the survey extending to $HK'$=20.5 mag for the
HDF and flanking fields.

\subsection{COMPLETENESS OF THE COMBINED OPTICAL-INFRARED CATALOGUE}

The final photometric catalogue of spheroidals was obtained by
matching the HST $I_{814}$-band and the ground-based IR {\tt
SExtractor} catalogues using the adopted magnitude limits of
$HK'<$19.5 mag and \ib$<$23.0 mag. In the final matched catalogue,
we retained the SExtractor `$m_{best}$' magnitudes but measured
$I-HK'$ colours within a fixed 3 arcsec diameter. This aperture
size, together with the fairly good seeing of the IR data, ensures
that when calculating colours we are looking at the same physical
region of the galaxy. Of the 818 sources in the matched catalogue,
266 systems were visually classified as spheroidals (defined to be
one of `E, E/S0, S0, or S0/a' in the MDS scheme) and 50 as compact
objects. Automated classifications result in 323 sources classed
as spheroidals (with no distinction between spheroidals and
compacts).

Clearly the joint selection by \ib and $HK'$ necessary to exploit
HST's morphological capabilities and establish optical-infrared
colours could lead to complications when interpreting $I-HK'$
colour distributions. As a major motivation for this study is to
identify as completely as possible the extent of any red tail in
the colour distribution, incompleteness caused by the various
magnitude limits is an important concern. Figure~5 shows that,
within the {\it optical, morphologically-selected} sample with
\ib$<$23 mag, virtually all of the $HK'<$19.5 mag sample can be
matched; only a small fraction (18/818=2.2\%) of red $I-HK'>$3.5-5
mag objects are missed. We return to the nature of these sources
in $\S$3.3.

\begin{figure}
\begin{center}
\leavevmode
\centerline{\psfig{file=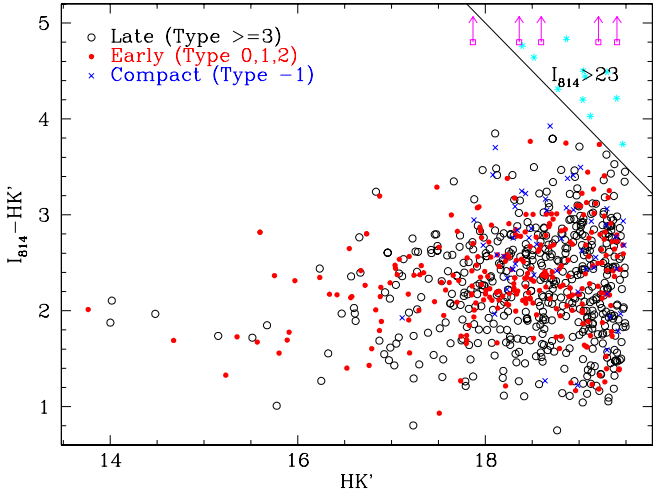,width=7cm,angle=-90}}
\end{center}
%\vspace{6cm}
\caption{\em Colour-magnitude distribution for the visually-classified
catalogue limited at $HK'$=19.5 mag and \ib=23 mag. Solid points correspond to
spheroidals, crosses to compacts and circles to the remaining spirals and
late type galaxies. The joint selection in \ib and $HK'$ implies a small
fraction ($<$3\%) of $HK'$-selected objects are not contained within
the HST sample. These objects are shown as stars and arrows as lower
limit when no detection was possible}
\label{fig:Figure5}
\end{figure}

\section{ANALYSIS}

\subsection{Strategy}

Our analysis is motivated by the two principal differences we might
expect between models where ellipticals underwent a strong initial
burst of activity with subsequent passive evolution (which we will
term the `monolithic collapse' model) and those associated with a
hierarchical assembly of ellipticals from the dynamical merger of
gas-rich disks (Baugh et al 1996). We recognise at the outset that
these models represent extreme alternatives with a continuum of
intermediate possibilities (c.f. Peacock et al 1998; Jimenez et al. 1998).
Our strategy in
this paper, however, will be to discuss our field elliptical data in
the context of the simplest models proposed to explain the star
formation history of distant {\it cluster} ellipticals (Ellis et al
1997, van Dokkum et al 1998). More elaborate analyses are reserved
until spectroscopic data is available for a large sample.

Firstly, in the monolithic collapse model, the comoving number density
of luminous ellipticals should be conserved to the formation redshift
(say, $z\simeq$3-5), whereas in hierarchical models we can expect some
decline in number density at moderate redshift depending on the
cosmological model and other structure formation parameters (Kauffmann
et al 1996, Kauffmann \& Charlot 1998a). Such a change in the
absolute number density would be difficult to convincingly detect
without spectroscopic data. The number of faint HST-identified
ellipticals has been discussed by Glazebrook et al (1995), Driver et
al (1995), Abraham et al (1996b) and Im et al (1996) with fairly
inconclusive results because of uncertainties arising from those in
the local luminosity function used to make predictions (Marzke et al
1998).

Secondly, there will be a redshift-dependent colour shift associated
with merger-driven star formation in the hierarchical models whereas,
for the monolithic case, the sources will follow the passive evolution
prediction. Kauffmann et al (1996) initially claimed that both
signatures would combine in the hierarchical picture to produce a
factor 3 reduction in the abundance of passively-evolving sources by
$z\simeq$1, but a more recent analysis (Kauffmann \& Charlot 1998b)
shows that the decline is dependent on the input parameters. For
example in a model with non-zero cosmological constant (the so-called
'$\Lambda$CDM'), little decline is expected until beyond z$\simeq$1.

In contemplating these hypotheses in the context of our HST data,
it must be remembered that although HST can be used very
effectively to isolate spheroidals morphologically to \ib=23 mag
(representing a considerable advance on earlier colour-selected
ground-based samples which could be contaminated by dusty later
types), in the case of merger models, the predictions will depend
critically on the time taken before a system becomes a
recognisable spheroidal. However, any hypothesis which postulates
a constant comoving number density of well-established spheroidals
is well suited for comparison with our data, the outcome being
important constraints on the past star formation history and
luminosity evolution.

\begin{figure}
\begin{center}
\leavevmode
\centerline{\psfig{file=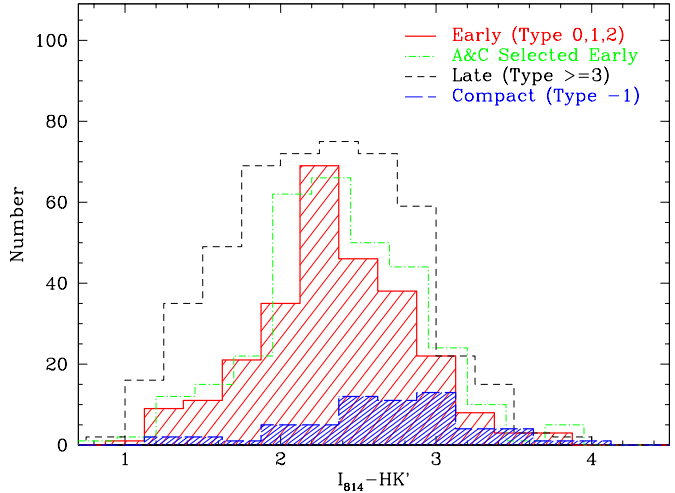,width=7cm,angle=-90}}
\end{center}
%\vspace{6cm}
\caption{\em Colour histograms for various morphological samples
within our adopted limits of $HK'<$19.5 and \ib=23. The solid line
refers to the visually-classified spheroidals (MDS types 0,1,2) and
the dashed-dot line to those spheroidals identified on the basis of their
asymmetry and concentration indices. The long dashed line refers to
visually-classed compacts (type=-1) and the short dashed line refers to the
remainder (types $\ge$3).}
\label{fig:Figure6}
\end{figure}

\subsection{Colour Distributions}

Figure 6 shows \ib-$HK'$ colour histograms for both the visual and
A/C-selected spheroidals alongside those for the compacts and the
remainder.  The automated and visual catalogues have nearly
identical colour distributions, confirming earlier tests on the
reliability of the automated classifier. In fact, the differences
between the automated and visually defined histograms are almost
completely attributable to the compact systems, which cannot be
segregated from other early-types on the basis of central
concentration. The colour histogram for compacts spans the range
seen for early-type galaxies, with a peak slightly redward of that
for visually classified early-types. It is clear from the
similarity between the colour histograms for visual and automated
classifications that contamination of spheroidals by compacts
(expected in the automated catalogue) does not pose a significant
uncertainty in determining the colour distribution.

The histogram of colours for late-type galaxies peaks at nearly
the same colour as that for the early-types, which at first seems
somewhat surprising. As we will later see, this is largely a
reflection of the very wide redshift range involved. However, the
distribution for spirals and later types is skewed toward the
blue, although redward of $I-HK'=$2.5 mag the shapes of the
distributions are similar (see also \S3.5 below).

\subsection{Single Burst Model Predictions}

Figure~7 compares the observed colour histograms with a range of model
predictions based on the GISSEL96 spectral synthesis code (Bruzual \&
Charlot 1996) for a range of star-formation histories. Observed and
predicted total counts for each of the models are also given in
Table~3.  At this stage we concentrate on `single burst' or
`monolithic collapse' models which conserve the comoving number
density at all epochs, and
defer discussion of alternative scenarios until \S3.5.

Our model predictions take into account the joint \ib and $HK'$
selection criteria for our sample, and are based on the
present-day optical E/S0 luminosity function (LF) from Pozzetti et
al. (1996), ie. a standard Schechter function with $\phi^\star=0.95\times
10^{-3}$ Mpc$^{-3}$, $M^{\star}_{b_{j}}=-20.87$ and a faint-end
slope of $\alpha=-0.48$. When making model predictions, this
luminosity function is tranformed into one appropriate for the \ib
photometric band via a single colour shift, resulting in
$M^{\star}_{I_{814}}=-23.12$. For comparison, we also show
predictions assuming a suitably transformed luminosity function
with a flat faint-end slope ($\alpha = -1$) and $\phi^\star=0.55\times
10^{-3}$Mpc$^{-3}$ as suggested by Marzke et at (1998). Throughout
this paper we adopt $H_0=50$ Km~s$^{-1}$Mpc$^{-1}$. Given the
elementary nature of the comparisons currently possible, and the
fact that the expected dispersion in $I-HK'$ from the present-day
colour-luminosity relation is minimal, we have avoided the
temptation to model a {\it distribution} of metallicities within
the galaxy population, preferring instead to explore the effects
of fixing the metallicity of the entire population to a single
value within a large range (40\%-250\% solar) in the simple
predictions discussed below.  Other variables in the single burst
hypothesis include the redshift of formation, $z_F$ (fixed at
$z_f=5$), the burst-duration (represented as a top hat function of
width 1.0 Gyr) and the cosmological parameters ($\Omega_M$ and
$\Omega_\Lambda$). As shown in Appendix A, the luminosity weighted
metallicities of the present sample are not biased strongly by the
limiting isophotes of the our observations, and fair comparisons
can be made using individual single-metallicity tracks over a
broad range of redshifts.

\begin{figure*}
\begin{center}
\leavevmode \centerline{\psfig{file=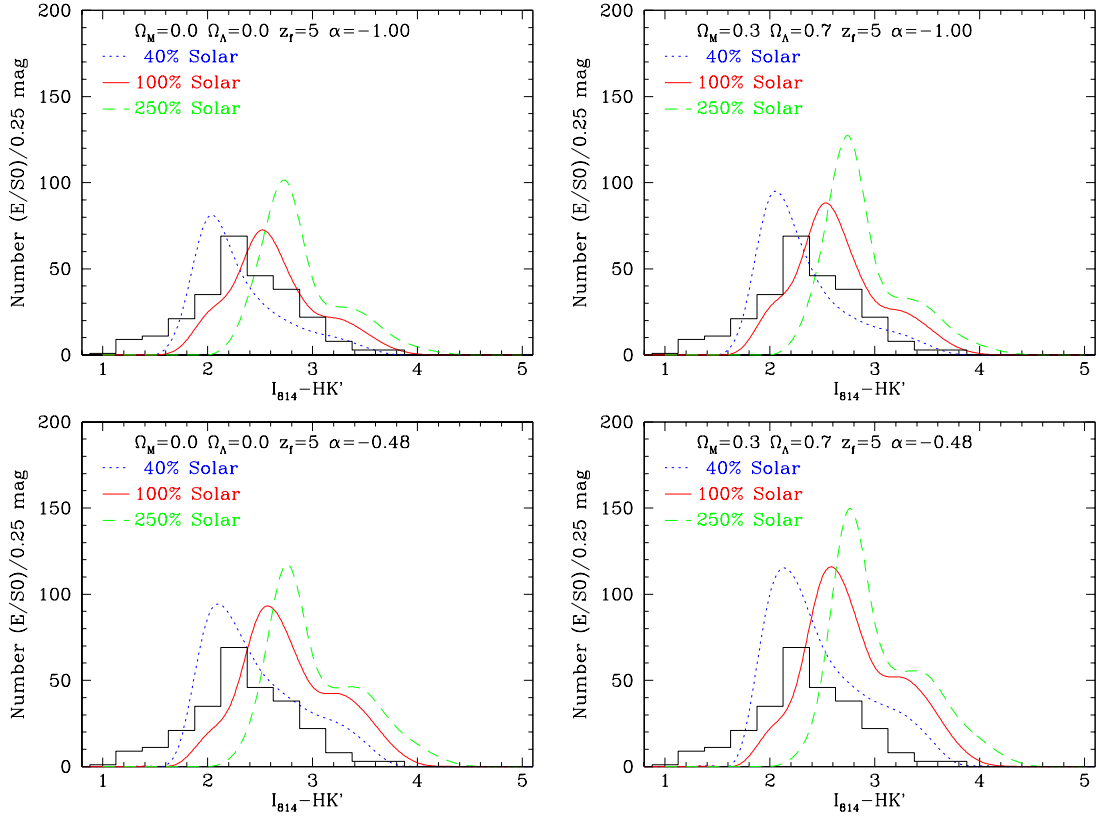,width=12cm,angle=-90}}
\end{center}
%\vspace{6cm}
\caption{\em The colour distribution of visually-classified spheroidals
with various single burst models (see legend) compared to the observed
number, represented by solid histogram (see text for further assumptions).}
\label{fig:Figure7}
\end{figure*}

Clearly the most important input parameter in the model
predictions shown in Figure~7 (summarized in Table~3) is the
metallicity. Although our spheroidals are almost exclusively
luminous ($>L^{\ast}$) galaxies which, in the context of
single-burst models would imply a metallicity of at least solar
(cf. Appendix A, and Arimoto et al 1997), here we will explore the
possibility that part of the colour distribution could arise from
a wider metallicity range than that found locally.

\subsubsection{A Deficit of Red Spheroidals}

The predicted colour distributions show a characteristic
asymmetry. This is caused by the blending of the $I_{814}-HK'$
K-correction and the passive bluing of systems to $z\simeq$1.5.  In
contrast, the observations reveal a clear excess of blue (\ib$-HK'<$2
mag) spheroidals not predicted by even the lowest metallicity
models. More significantly, solar and super-solar metallicity models
over-predict the extent of the red tail in the colour
distribution. Both discrepancies are consistent with recent
star-formation in our sample of faint spheroidals. In order to
quantify these discrepancies, we performed a Kolmogorov-Smirnov test
(K-S) to check whether the observed spheroidal colours could be drawn
from any of the model distributions (allowing a measurement error
$\sigma_{I-HK'}=0.25$ mag). In all cases the observed distribution
differs from the models at a confidence level higher than
99.99\%. Evidently monolithic collapse models with constant co-moving
density fail to reproduce the colour distribution of high redshift
spheroidals.

It will be convenient in the following to quantify the strength of
the red tail in the colour distribution by defining a ``red
fraction excess'', shown in Table~3, constructed as the ratio of the
predicted number of early type systems with $I_{814}-HK' > 3.0$
mag to the observed number. The statistical uncertainties on the
red fraction excess in this table are based on 500 bootstrap resamplings
of the original catalogue, each realization of which was subjected
to the same selection criteria applied to the original data.

As discussed by many authors (Glazebrook et al 1995, Marzke et al
1998), the absolute numbers depends sensitively on the normalisation
and shape of the local LF. Table 3 includes a summary for the range in
LF parameters discussed earlier. For a declining faint-end slope
($\alpha=-0.48$) and solar-and-above metallicity, the red fraction
excess is more than 5 times that observed.  Adopting a metallicity
substantially below solar results in closer agreement but assuming
such metallicities for the entire population may be unreasonable given
local values (see Appendix, and Arimoto et al 1997). Even so, the red
fraction excess is only reduced from $\sim 8$ to $\sim 3$ if the adopted
metallicity is varied between 250\% solar and 40\% solar. Models with
a flat faint-end slope ($\alpha=-1$) improve the agreement further and
in the very lowest metallicity model with $\Omega_M=0$ there is almost
no deficit.

In Table~3 we have also included models with $\Omega_{\Lambda}>0$ for
both slopes of the LF. The effect of $\Omega_{\Lambda}$ is to increase
the apparent deficit of red spheroidals (because of the rapid increase
in the differential volume element with redshift for
$\Omega_{\Lambda}>0$ cosmologies at $z<1$).

Although models where the redshift of the initial burst, $z_F$, is
reduced to $z=3$ are not shown in the table, these do not result
in significant changes in the above discussions. We conclude that
we cannot reconcile the number of galaxies in the red end of the
observed colour histogram to the corresponding predictions of a
constant co-moving density high-redshift monolithic collapse
model. Alternative scenarios, which may explain the relatively
blue colours of some observed spheroidals, will be considered in
\S3.5.

\subsubsection{A Declining Number of High Redshift Spheroidals?}

While monolithic collapse models fail to reproduce the observed colour
distributions, Table~3 indicates that the overall number is in
reasonable agreement. Specifically, for a low $\Omega_M$ and
$\Omega_\Lambda$=0, the Marzke et al. luminosity function and
luminosity-weighted metallicities of solar and above, we see no
significant evolution in the space density of spheroidals.  For the
currently popular spatially-flat universe with low $\Omega_M$ and high
$\Omega_\Lambda$ (Perlmutter et al 1999), the data imply a deficit of
no more than 30\%. Stronger evolution ($\sim$ 60\% decline) would
occur if we adopted the Pozetti et al. luminosity function. We
therefore conclude that the colour offset described earlier is more
likely the result of star-formation activity in well-formed
spheroidals at high redshifts rather than evidence for evolution in
their space density.

At this point, we return to the nature of those 18 sources
identified in the infrared images which are fainter than \ib=23
mag. Although they could formally be included in the colour
distributions, they are too faint in the WFPC-2 images for
reliable morphological classification. A montage of these sources
is shown in Figure~8. At most 3 of the 18 are {\em possible}
spheroidals. As such, their addition to the colour distribution
would have a negligible impact on conclusions drawn from Figure~7.

\begin{figure*}
\begin{center}
\leavevmode
\centerline{\psfig{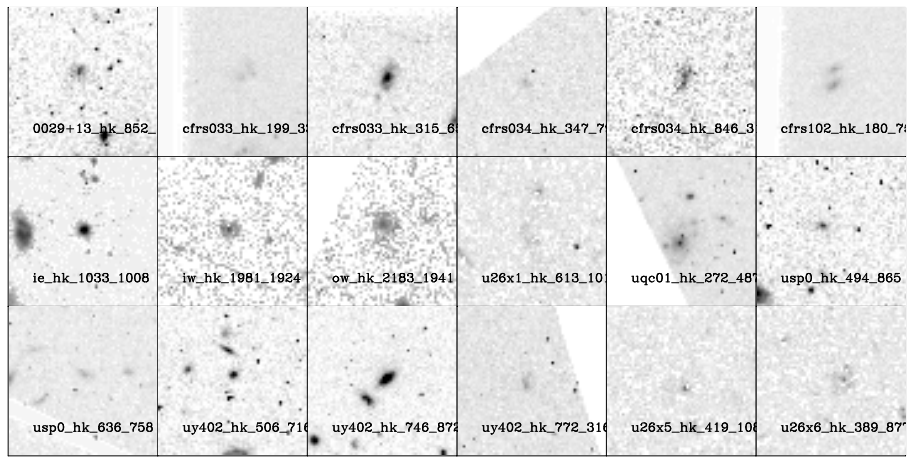}}
\end{center}
%\vspace{6cm}
\caption{\em A montage of WFPC-2 images for those red sources contained
within the $HK'<$19.5 sample but for which \ib$>$23.}
\label{fig:Figure8}
\end{figure*}

\subsection{Constraints from Redshift Distributions}

\begin{figure}
\begin{center}
\leavevmode
\centerline{\psfig{file=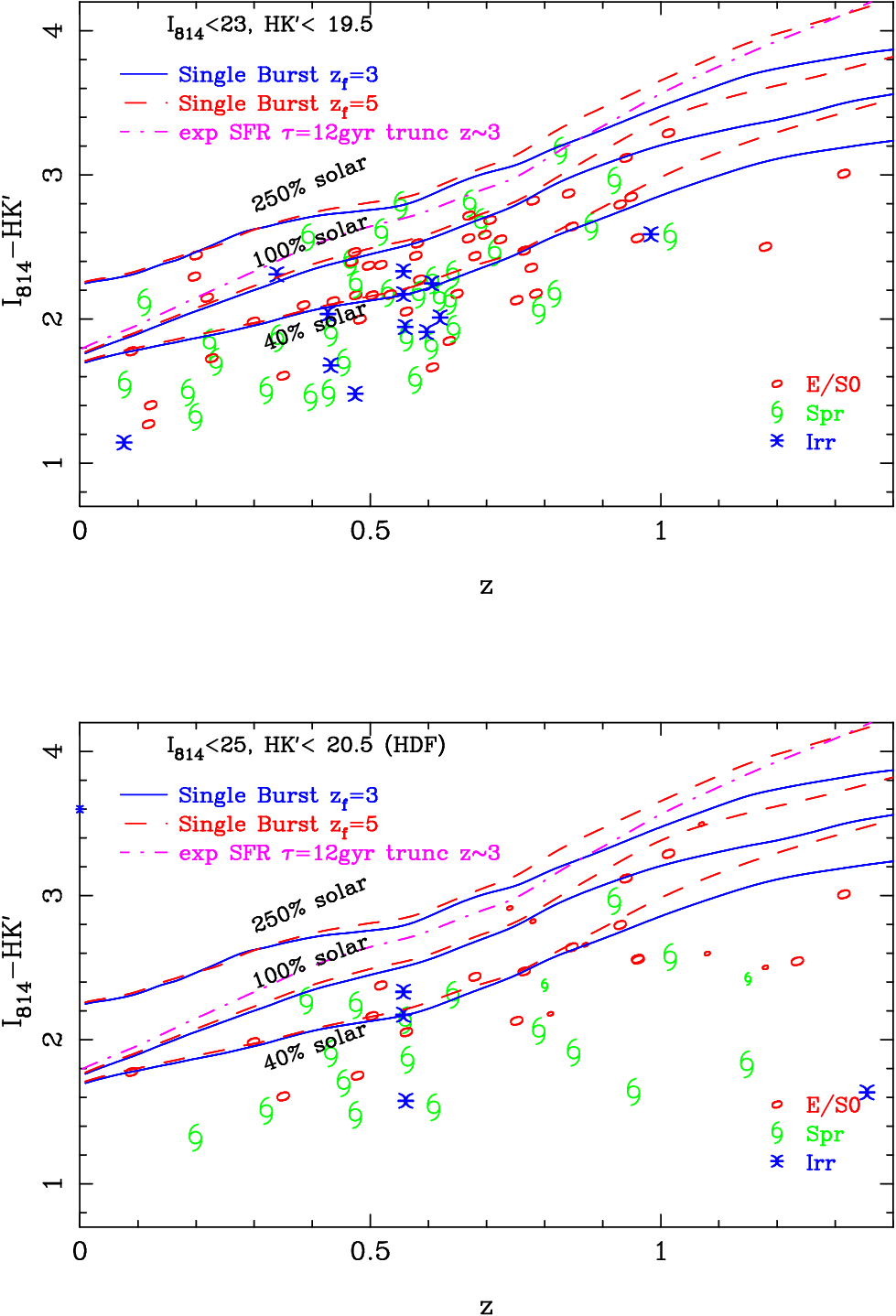,width=9cm,angle=0}}
\end{center}
%\vspace{6cm}
\caption{\em The colour-redshift relation for that subset of the
$HK'<$19.5 mag sample with published spectroscopic data (large
symbols), and for a deeper $HK'<$20.5 mag sample within the HDF for
which photometric redshift data (smaller symbols) is available (Wang
et al 1998). Symbols refer to visual classifications. The solid lines
represents a passively evolving elliptical (single burst of duration
0.1 Gyr) at $z_F$=5 for different metallicities. Dashed lines show the
corresponding models for $z_F$=3. Dot-dashed lines represent an
exponential star formation with e-folding time $\tau=12$Gyr, truncated
at $z=$3, with solar metallicity}
\label{fig:Figure9}
\end{figure}

While our principal conclusions are based on the enlarged size of
our HST sample and addition of infrared photometry compared to
earlier work, it is interesting to consider what can be learnt
from the (incomplete) redshift data currently available for our
sample. We have collated the published spectroscopic redshift data
from the CFRS/LDSS surveys (Brinchmann et al 1998), the MDS survey
(Glazebrook et al 1998) and the HDF and its flanking fields
(tabulated by Cowie 1997) and matched these with our
$HK'$-selected sample. In total, 97 of our galaxies have published
redshifts. As the bulk of these surveys were themselves
magnitude-limited in $I$, the magnitude and colour distribution of
this subset should be representative of that for an appropriate
subset of our primary photometric sample.

Figure~9 (upper panel) shows the colour-redshift relation for the
97 objects in the subsample with redshift information. Also shown
are evolutionary predictions based on spectral synthesis models
adopting ranges in metallicity and star-formation history as
before. The 47 spheroidals in this set are clearly redder at a
given redshift than their spiral and later-type counterparts and
span a wide redshift range with median value
$\overline{z}\simeq$0.7. However, as Schade et al (1998) discussed
in the context of \vb-\ib colours for their smaller sample of HST
field galaxies, there is some overlap between the classes at a
given redshift. The colour scatter for morphological spheroidals
appears somewhat larger than the $\sim 0.2$ mag dispersion
expected from slope of the infrared-optical colour-luminosity
relations for early-type systems\footnote{Note however that the
slope of the infrared-optical colour-magnitude relation for
early-type systems is rather uncertain, particularly for $I-K$. On
the basis of quite strongly model-dependent conversions based on
the cluster $V-K$ data of Bower et al. (1992), Peletier \& de
Grijs (1998) obtain a slope of $-0.0438 \pm 0.0041$ for the $I-K$
slope of local early-type systems, using the models presented in
Vazdekis et al. (1996).}.

In the specific case of the HDF, it is interesting to exploit the
increased depth of both the $HK'$ data and the HST optical
morphologies (Abraham et al 1996a) as well as to consider the
abundant photometric redshift estimates.  For this purpose we
constructed a 19.5$<HK'<$20.5 mag extension to our HDF sample,
with morphological classifications from the deep (\ib $< 25$ mag)
morphological catalogue of van den Bergh (1996). For this sample
we can take advantage of the apparently rather good precision in
photometric redshift estimates for early-type galaxies (Connolly
et al 1997, Wang et al 1998)\footnote{We note that the good
accuracy in photometric redshifts for these galaxies appears to be
due to the presence of strong continuum features which are
well-explored with the addition of the Kitt Peak JHK photometry to
the HDF filter bands.}. By going deeper in the HDF we extend our
sample by 20 objects, of which 8 are classed visually as E/S0s
(none are compact). Adding this extension to those HDF galaxies
already in our catalog, the corresponding numbers with $HK'<20.5$
mag; $I_{814} < 25$ mag become 50 and 26 respectively. The
colour-redshift relation for the combined HDF sample is shown in
Figure~9 (lower panel). For the same sample, Figure~10 shows the
colour-magnitude diagram of the visually classified E/S0s. The
inset shows the colour histogram and the arrow indicates the peak
in the distribution for the primary sample.

As expected, the peak of the HDF colour histogram in Figure~10
lies redward of the colour histogram for our entire sample (by
$\sim 0.2$ mag). But the redshift data in Figure~9 makes it clear
that this peak is still substantially bluer than expected for the
simple monolithic collapse model at solar metallicity. For the 26
HDF spheroidals, spectroscopic redshifts are available for 19, the
rest being photometric. Importantly, the
spectroscopically-confirmed galaxies include 3 ellipticals beyond
z$\simeq$0.9, all of which are substantially bluer than the
passive evolution predictions. While based on small numbers of
galaxies, the figure lends strong support to the conclusions of
the previous subsection, particularly when it is realised there is
an in-built bias in favour of photometric redshifts matching the
passively-evolving spectral energy distributions.

These conclusions based on the HDF are consistent with those of
Zepf (1997) and Barger et al (1998) who analysed optical-infrared
colours of much fainter sources without taking into account
morphological and redshift information.

\begin{figure*}
\begin{center}
\leavevmode
\centerline{\psfig{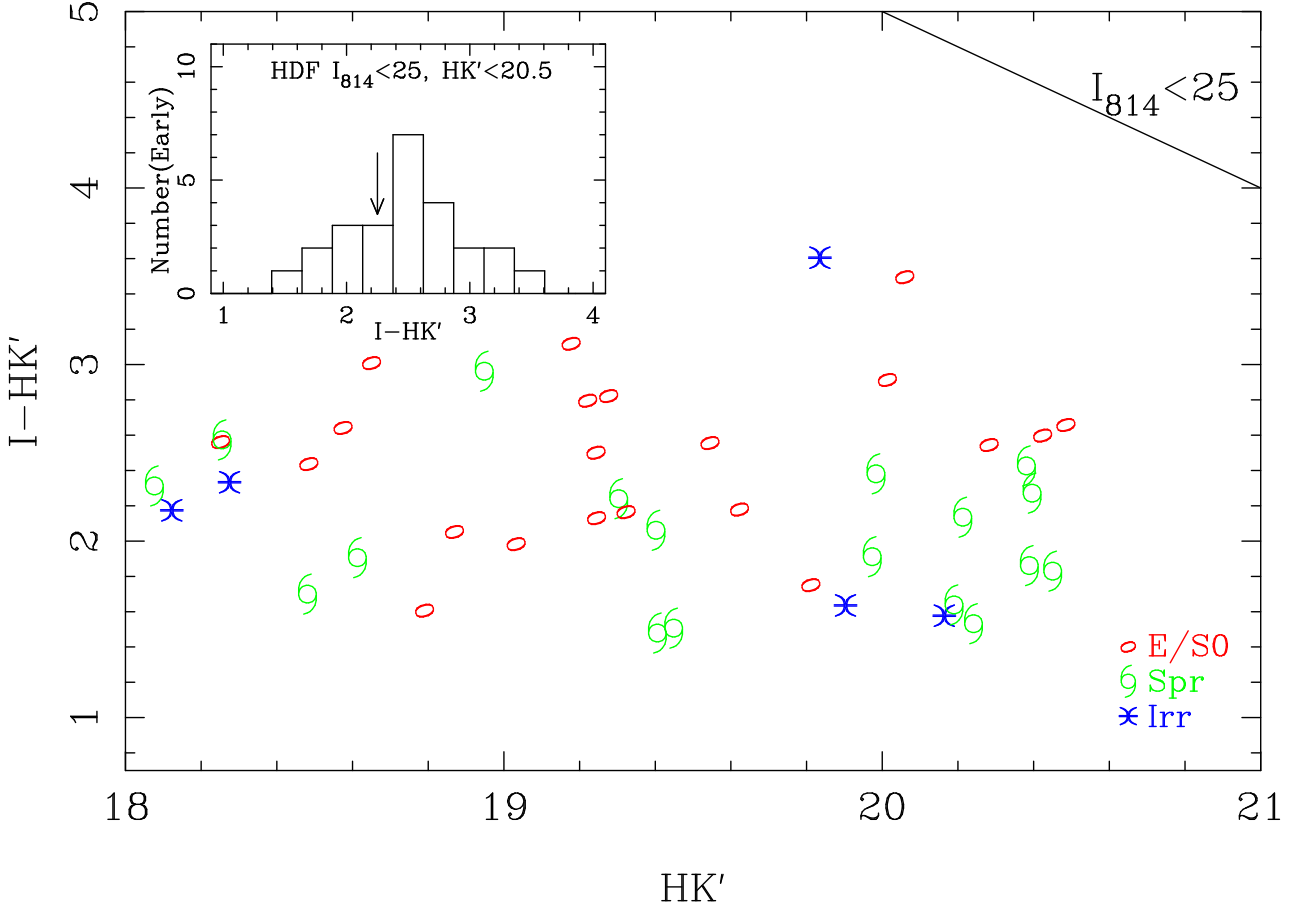}}
\end{center}
%\vspace{6cm}
\caption{\em The colour-magnitude diagram of visually-classed
objects within the deeper sample possible solely for the HDF
(\ib$<23$, $HK'<$20.5). The inset histogram shows the $I_{814}-HK'$ colour
distribution for visually-classed spheroidal within the deeper
sample. The arrow shows the peak of the distribution in the
primary sample. Symbols are as for Figure 9}
\label{fig:Figure10}
\end{figure*}

\subsection{Alternative Star Formation Histories}

The single burst models ruled out by the colours of speroidals in
the previous sections are idealised representations of spheroidal
history. We now consider alternative histories which could be more
consistent with our various datasets.

At its most fundamental level, the deficit of red spheroidals at faint
limits appears to eliminate models with very short epochs of star
formation at high redshifts followed by long quiescent periods. In the
context of modelling distant red radio galaxies, Peacock et al (1998)
have shows that models with continuous star formation truncated at
later times avoid the peak luminosities associated with initial bursts
and can produce a significant bluing at redshifts where the red tail
would otherwise be seen\footnote{On the basis of model predictions
used to calculate the density of post-starburst ``$H\delta$ strong''
systems seen in clusters, Couch \& Sharples (1986), Barger et
al. (1996) and Abraham et al. (1996c) note that a sharp truncation in
the star-formation rate of actively star-forming systems results in a
synchronization of optical colours with those expected of early type
galaxies after only $\sim 1.5$ Gyr, so it is not too surprising that
truncated star formation histories and monolithic collapse models
predict similar colours {\em provided} the epoch of observation is
several Gyr after the period of truncation.}. In an important recent
paper, Jimenez et al. (1998) argue that monolithic models have been
rejected prematurely by some authors: only {\em extreme} scenarios
with very short duration bursts (eg. $10^7$ Myr) of star-formation
followed by absolute quiescence can be ruled out, while bursts with
fairly low-levels of extended star-formation activity may still be
compatible with the observed data. On the basis of a simple
one-dimension chemo-dynamical model for the evolution of spheroids,
these authors predict that the integrated star-formation history of
early-type galaxies should resemble a quasi-monolithic collapse
model. A concrete prediction of this model is that the bulk of the
star-formation in high-$z$ spheroids is occuring near their cores, an
effect that appears to have been seen (Abraham et al. 1999), and which
may be responsible for the blue colours of some ellipticals in our
sample.

\begin{figure*}
\begin{center}
\leavevmode \centerline{\psfig{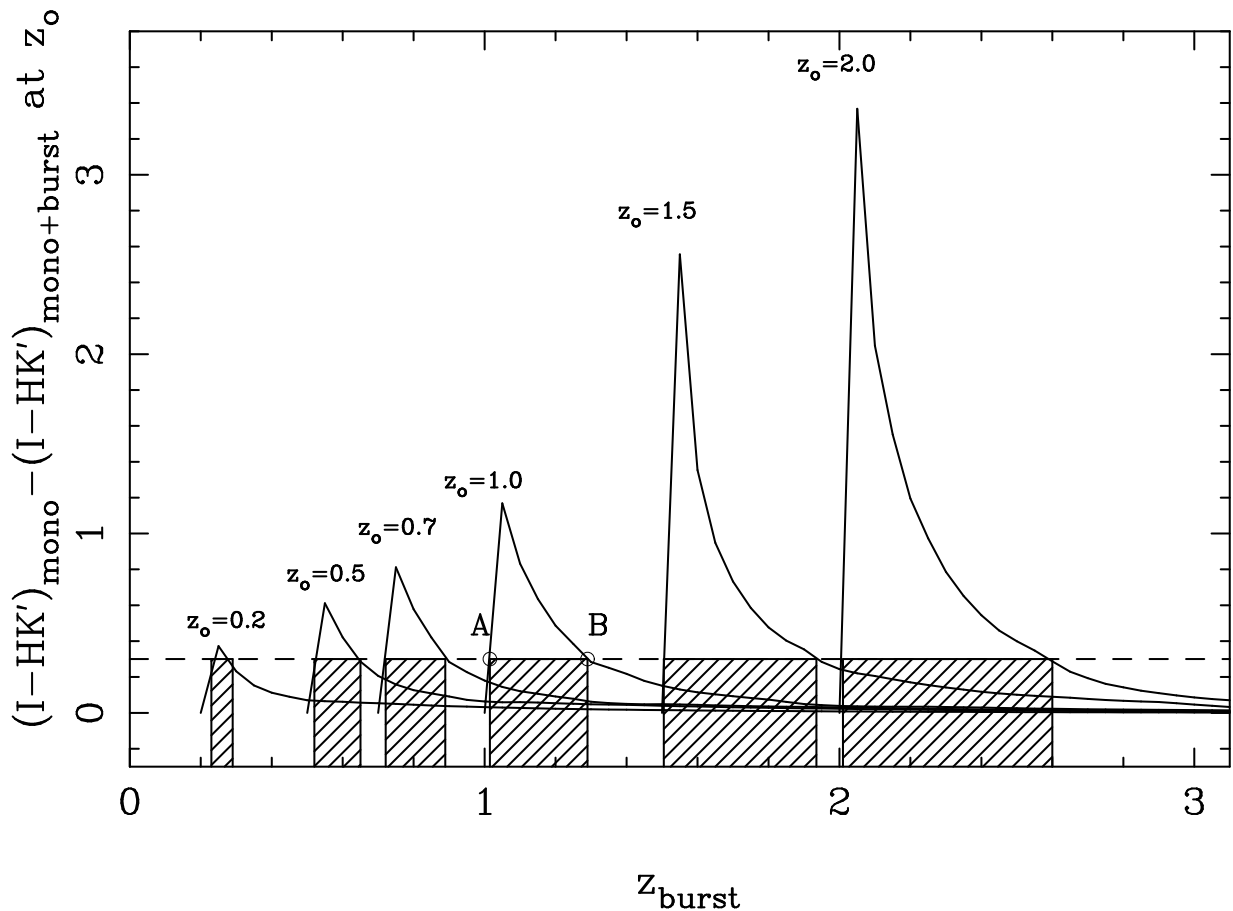}}
\end{center}
%\vspace{6cm}
\caption{\em The colour difference between a 1.0 Gyr single burst
and the same model plus a short burst of star formation as a
function of the redshift, $z=z_{burst}$, where the burst is added
is show for a series of redshift of observation ($z_0 =
0.2,0.5,0.7,1.0,1.5 \mbox{~and~} 2.0$). The horizontal dashed line
represent $\delta_{colour}=0.3$ and the shaded area shows the
range in $z_{burst}$ where $\delta_{color} \ge 0.3$ for each
$z_o$. }
\label{fig:Figure11}
\end{figure*}

In the light of the above, it is therefore interesting to
calculate the amount of recent star formation which must be added
to a single burst model in order for the model to match the
observed \ihk colours. We seek to determine the timescales over
which such a system is bluer than the passive evolution model by
at least $\delta_{colour} = (I_{814}-HK')_{model} -
(I_{814}-HK')_{obs} \simeq 0.3$ mag, i.e. consistent with the
typical colour offset from the solar metallicity tracks shown in
Figure 9. To model this we added a short burst of star-formation
(of duration 0.1 Gyr, forming 15\% of the stellar mass) occurring
at $z=z_{burst}$ to an underlying solar metallicity monolithic
collapse model at a given redshift using the population synthesis
models of Bruzual \& Charlot (1996). We then computed the redshift
range over which the burst at $z_{burst}$ results in
$\delta_{color}>0.3$ for a given redshift of observation, $z_o$.

The result of this exercise is shown in Figure~11 for several
redshifts of observation $z_0$. The range in redshift space where a
15\% burst produces a blueing $\delta_{colour} \ge 0.3$ is shown as
the shaded area. (For example, a galaxy observed at redshift $z_o=1$
would be at least 0.3 mag bluer than the monolithic collapse models
between points A and B in the figure.  At $z_o=0.5$ a burst would have
to occur at $0.5 <z_{burst}<0.65$, while for $z_o=0.7$ it would have
to occur in the range $z_{burst}\sim 0.7-0.9$. Since almost all points
in Figure~9 lie blueward of the solar metallicity model tracks, and
since galaxies in our sample lying in the redshift range $0<z<1$ have
``memory'' of bursts over $\delta z \sim 0.1-0.2$, it seems improbable
that moderate intensity (ie. 15\% mass) single burst events can
explain the colour offsets shown in Figure~9. It seems more probable
that the duty-cycle of star-formation is more extended, indicative of
either a low level of roughly continuous star-formation underlying the
old stellar populations, or perhaps of a succession of lower mass
bursts.

The preceeding analysis indicates the extent to which a modest
``polluting'' star-forming population superposed on a dominant old
stellar population reconciles our observations with traditional
monolithic collapse scenarios. In contrast to this, it is interesting
to consider what sort of hierarchical formation models may also be
consistent with the present data.  The rather mild density evolution
in our sample is in sharp disagreement with the predicted factor of
three decline in the abundance of spheroidals at $z \sim 1$, based on
the present generation of ``semi-analytical'' models in high-density,
matter-dominated cosmologies (eg. Kauffmann \& Charlot
1998a). However, in a flat $\Lambda$-CDM model (with $\Omega_M=0.3$,
$\Omega_\Lambda=0.7$) the decline in the abundance of spheroidals is
only 30\% at $z \sim 1$ (Kauffmann \& Charlot 1998b), and may be
consistent with the present observations. In the latter model the
oft-quoted factor of three decline in the space density of ellipticals
occurs at $z=2$ instead of $z=1$, so future work extending the present
sample to higher redshifts may allow us to distinguish between
``extended'' monolithic collapse scenarios and $\Lambda$-CDM
hierarchical models.  However, more detailed modelling in this
particular case, e.g. of the colours distribution, is beyond the scope
of this paper. If star-formation in extended monolithic-collapse
scenarios is centrally concentrated (as suggested by the simple
one-dimensional models of Jimenez et al.  1998), the two scenarios may
also be distinguishable at lower redshifts on the basis of {\em
resolved} colour distributions (Menanteau et al. 1999, in
preparation).

\section{CONCLUSIONS}

We have constructed a new catalogue of $\simeq$300 faint field
spheroidal galaxies using HST images for morphology to a limit of
\ib=23 mag. Follow-up infrared photometry has enabled us to
consider the optical-infrared colour distribution of an infrared
magnitude limited sample. We have modelled expected colours using
various star formation histories, metallicities and cosmologies.
For a limited subset, spectroscopic redshift data is available and
within the HDF it is possible to construct a deeper catalogue.

Our main results can be summarised as follows:

$\bullet$ There is little evidence for strong evolution in the
space density of luminous field spheroidal systems out to $z \sim
1$. Within the uncertainties introduced by poorly constrained
values for the local spheroidal luminosity function and by
$\Omega_M$ and $\Omega_\Lambda$, our data is consistent with no
evolution, or with modest evolution (at the level of $\sim 30\%$)
in terms of a decline in the space density of spheroidals by $z
\sim 1$.

$\bullet$ Although we detect little evidence for strong evolution in
the space density of luminous spheroidals, we find a marked deficit in
the number of red spheroidals compared to predictions where the bulk
of star formation was completed prior to $z\simeq$3.

$\bullet$ Where redshift data is available it suggests that the
duty cycle for star-formation in high-redshift spheroidals is
indicative of a low level of extended star-formation underlying a
passively evolving population, rather than of relic star-formation
following from a massive burst episode.

$\bullet$ The apparently mild density evolution and blue colours
of high-redshift spheroidals in the present sample are consistent
with the predictions of ``extended'' monolithic collapse
scenarios, in which the existing star-formation pollutes the
colours of a dominant, underlying old stellar population.  The
data is {\em not} consistent with the predictions of
semi-analytical hierarchical models in high-density,
matter-dominated cosmologies. However, the observed weak density
evolution may be consistent with the predictions of hierarchical
$\Lambda$-CDM models (Kauffmann \& Charlot 1998b).

Spectroscopic redshifts for a complete sub-sample of our catalogue
will enable models for the star-formation history of high redshift
spheroidals to be rigorously tested, eliminating some of the
ambiguities present in the current analysis.

\section*{ACKNOWLEDGMENTS}

FM would like to thank PPARC and Fundanci\'on Andes for financial
support. RGA acknowledges support from a PPARC Advanced
Fellowship. We acknowedge substantial comments from an anonymous
referee which transformed an earlier version of this paper. We
thank Simon Lilly and Chuck Steidel for valuable input.

\begin{table*}
\centering
\begin{minipage}{140mm}
\caption{The HST Sample}
\begin{tabular}{@{}lrrrrrrrr}
\hline
Field     & $I_{814}$ & $V_{606}$ & $H$+$K'$  & FWHM
&$\alpha$(J2000) & $\delta$(J2000) & N$_z$ & N$_{E/S0}$\\
\hline\hline
0029+13  &   6300 &   3300 &  3900 & 0.6" &  00:29:06.2 & 13:08:12.9   &    ...  &    ...  \\
0144+2   &   4200 &   ...  &  2340 & 0.8" &  01:44:10.6 & 02:17:51.2   &    ...  &    ...  \\
0939+41  &   4600 &   5400 &  3120 & 0.9" &  09:39:33.3 & 41:32:47.9   &    ...  &    ...  \\
1210+39  &   4899 &   3999 &  3120 & 0.6" &  12:10:33.3 & 39:29:01.6   &    ...  &    ...  \\
1404+43  &   8700 &   5800 &  3120 & 0.7" &  14:04:29.3 & 43:19:15.2   &    ...  &    ...  \\
East     &   5300 &   ...  & 10985 & 0.8" &  12:37:02.0 & 62:12:23.4   &    ...  &    ...  \\
HDF      & 123600 & 109050 & 10530 & 0.8" &  12:36:47.5 & 62:13:04.4 &30 &18  \\
NEast    &   2500 &   ...  & 10595 & 0.8" &  12:37:03.7 & 62:15:13.0   &    ...  &    ...  \\
NWeast   &   2500 &   ...  & 10985 & 0.8" &  12:36:49.4 & 62:15:53.8   &    ...  &    ...  \\
OutEast  &   3000 &   ...  & 10985 & 0.8" &  12:37:16.2 & 62:11:42.5   &    ...  &    ...  \\
OutWest  &   2500 &   ...  & 10270 & 0.8" &  12:36:19.3 & 62:14:25.5   &    ...  &    ...  \\
SEast    &   2500 &   ...  & 10465 & 0.8" &  12:36:46.2 & 62:10:14.6   &    ...  &    ...  \\
SWest    &   2500 &   ...  & 10395 & 0.8" &  12:36:32.0 & 62:10:55.3  &    ...  &    ...  \\
West     &   5300 &   ...  & 11375 & 0.8" &  12:36:33.6 & 62:13:44.9   &    ...  &    ...  \\
cfrs031  &   6700 &   ...  &  3120 & 0.8" &  03:02:57.4 & 00:06:04.9 &8 &4  \\
cfrs033  &   6700 &   ...  &  3120 & 0.7" &  03:02:33.0 & 00:05:55.2 &5 &1  \\
cfrs034  &   6400 &   ...  &  3120 & 0.8" &  03:02:49.8 & 00:13:09.2 &5 &1  \\
cfrs035  &   6400 &   ...  &  3120 & 1.0" &  03:02:40.8 & 00:12:34.3 &2 &1  \\
cfrs101  &   6700 &   ...  &  3120 & 0.8" &  10:00:23.2 & 25:12:45.1 &7 &2  \\
cfrs102  &   6700 &   ...  &  3120 & 0.7" &  10:00:36.5 & 25:12:40.1 &9 &6  \\
cfrs103  &   5302 &   ...  &  3120 & 0.6" &  10:00:46.7 & 25:12:26.3 &12 &7  \\
u26x1    &   4400 &   2800 &  4420 & 0.8" &  14:15:20.1 & 52:02:49.9   &    ...  &    ...  \\
u26x2    &   4400 &   2800 &  3120 & 0.7" &  14:15:13.7 & 52:01:39.6   &    ...  &    ...  \\
u26x4    &   4400 &   ...  &  3120 & 0.7" &  14:18:03.2 & 52:32:10.4   &    ...  &    ...  \\
u26x5    &   4400 &   ...  &  3315 & 0.7" &  14:17:56.7 & 52:31:00.6    &    ...  &    ...  \\
u26x6    &   4400 &   ...  &  3120 & 0.5" &  14:17:50.3 & 52:29:50.8    &    ...  &    ...  \\
u26x7    &   4400 &   ...  &  3120 & 0.7" &  14:17:36.9 & 52:27:31.2    &    ...  &    ...  \\
u2ay0    &  25200 &  24399 &  4160 & 1.4" &  14:17:42.7 & 52:28:31.3   &    ...  &    ...  \\
u2iy     &   7400 &   ...  &  3120 & 0.6" &  14:17:43.1 & 52:30:23.3  &    ...  &    ...  \\
u3d3     &   4200 &   3300 &  3120 & 1.3" &  20:29:39.6 & 52:39:22.3   &    ...  &    ...  \\
ubi1     &   6300 &   3300 &  2600 & 0.6" &  01:10:00.7 &-02:27:22.2 &5 &1  \\
uci1     &  10800 &   4800 &  3900 & 0.4" &  01:24:41.6 & 03:51:24.2   &    ...  &    ...  \\
ueh0     &  12600 &   8700 &  3900 & 0.5" &  00:53:24.1 & 12:34:01.9 &4 &2  \\
uim0     &  11800 &   ...  &  4090 & 0.6" &  03:55:31.4 & 09:43:31.9 &5 &1   \\
umd4     &   9600 &   2400 &  3900 & 0.7" &  21:51:06.9 & 28:59:57.0   &    ...  &    ...  \\
uop0     &   4200 &   7200 &  3120 & 0.9" &  07:50:47.9 & 14:40:39.1   &    ...  &    ...  \\
uqc01    &   7200 &   7800 &  3120 & 1.0" &  18:07:07.3 & 45:44:33.5   &    ...  &    ...  \\
usa0     &   6300 &   5400 &  3120 & 0.7" &  17:12:24.0 & 33:35:51.2   &    ...  &    ...  \\
usa1     &   6300 &   5400 &  3900 & 0.5" &  17:12:24.7 & 33:36:03.0   &    ...  &    ...  \\
usp0     &   4200 &   3300 &  3120 & 0.9" &  08:54:16.9 & 20:03:35.5   &    ...  &    ...  \\
ust0     &  23100 &  16500 &  3120 & 0.9" &  10:05:45.8 &-07:41:21.0   &    ...  &    ...  \\
ut21     &  11999 &   3600 &  3900 & 0.8" &  16:01:13.2 & 05:36:03.2   &    ...  &    ...  \\
ux40     &   7500 &   3300 &  3900 & 0.6" &  15:19:40.3 & 23:52:09.9 &5 &3  \\
ux41     &   6000 &   3300 &  3120 & 0.8" &  15:19:54.6 & 23:44:57.8   &    ...  &    ...  \\
uy402    &   5400 &   1400 &  3120 & 0.8" &  14:35:16.0 & 24:58:59.9   &    ...  &    ...  \\
uzk0     &   8700 &   8400 &  3120 & 0.9" &  12:11:12.3 & 39:27:04.8   &    ...  &    ...  \\
uzp0     &   6300 &   3300 &  3120 & 0.8" &  11:50:28.9 & 28:48:35.2   &    ...  &    ...  \\
uzx05    &   4700 &   2600 &  3120 & 0.7" &  12:30:52.7 & 12:18:52.5  &    ...  &    ...  \\
\hline
\end{tabular}
\end{minipage}
\end{table*}

\begin{table*}
\centering
\begin{minipage}{140mm}
\caption{Log of Observations}
\begin{tabular}{@{}lllll}
\hline
Program Date     & Fields Observed & $\langle$Seeing$\rangle$ & Telescope \\
\hline\hline
Feb   6-8 1996 & HDF           & ~1.0" & UH-2.2m \\
Apr   5-8 1996 & HDF           & ~1.0" & CFHT    \\
Apr 17-22 1996 & 8 HDF-FF      & ~0.8" & UH-2.2m \\
Aug 20-24 1997 & 21 MDS fields & ~0.8" & UH-2.2m \\
Jan 18-20 1998 & 23 MDS fields & ~0.8" & UH-2.2m \\
Feb 12-13 1998 & 11 MDS fileds & ~0.7" & UH-2.2m \\
\hline
\end{tabular}
\end{minipage}
\end{table*}

\begin{table*}

\centering
\begin{minipage}{140mm}
\footnotetext[1]{Single Burst Model with 1.0 Gyr duration and $z_F=5$.}
\footnotetext[2]{${N_{total, model}}/{N_{early,observed}}$, based on
visual classifications with 1$\sigma$ upper and lower limits.}
\footnotetext[3]{${N_{early, model}^{I_{814}-HK'>
3.0}}/{N_{early,observed}^{I_{814}-HK'>3.0}}$, based on visual
classifications with 1$\sigma$ upper and lower limits.}
\caption{Absolute Numbers}
\begin{tabular}{@{}lllll}
\hline
Data Set   & $N_{early}$ & $N_{early}^{I_{814}-HK'> 3.0}$ & Total
Excess$^{b}$ & Red Fraction Excess$^{c}$ \\
\hline\hline
\sc{Observational data (with bootstrapped errors)}  \\ \\
Visually Classified            & $266 \pm 13$ & $26 \pm 5$ \\
Visually Classified + Compacts & $316 \pm 14$ & $40 \pm 5$ \\
A/C classified                 & $323 \pm 13$ & $40 \pm 6$ \\ \\

\sc{Pozzetti et al LF$^{a}$ ($\alpha=-0.48$)}  \\ \\

~40\% Solar $\Omega_M=0.0$ $\Omega_{\Lambda}=0.0$  & 357.17 &  72.63 & $   1.34_{ 1.28}^{ 1.41}$  & $   2.79_{ 2.34}^{ 3.46}$  \\
100\% Solar $\Omega_M=0.0$ $\Omega_{\Lambda}=0.0$  & 366.57 & 135.26 & $   1.38_{ 1.31}^{ 1.45}$  & $   5.20_{ 4.36}^{ 6.44}$  \\
250\% Solar $\Omega_M=0.0$ $\Omega_{\Lambda}=0.0$  & 372.85 & 198.09 & $   1.40_{ 1.34}^{ 1.47}$  & $   7.62_{ 6.39}^{ 9.43}$  \\\\

~40\% Solar $\Omega_M=0.3$ $\Omega_{\Lambda}=0.7$  & 433.94 &  90.37 & $   1.63_{ 1.56}^{ 1.72}$  & $   3.48_{ 2.92}^{ 4.30}$  \\
100\% Solar $\Omega_M=0.3$ $\Omega_{\Lambda}=0.7$  & 449.58 & 168.27 & $   1.69_{ 1.61}^{ 1.78}$  & $   6.47_{ 5.43}^{ 8.01}$  \\
250\% Solar $\Omega_M=0.3$ $\Omega_{\Lambda}=0.7$  & 458.74 & 247.12 & $   1.72_{ 1.64}^{ 1.81}$  & $   9.50_{ 7.97}^{11.77}$  \\\\

\sc{Marzke et al LF$^{a}$ ($\alpha=-1.00$)}       \\ \\

~40\% Solar $\Omega_M=0.0$ $\Omega_{\Lambda}=0.0$  & 240.29 &  28.56 & $   0.90_{ 0.86}^{ 0.95}$  & $   1.10_{ 0.92}^{ 1.36}$  \\
100\% Solar $\Omega_M=0.0$ $\Omega_{\Lambda}=0.0$  & 259.81 &  65.18 & $   0.98_{ 0.93}^{ 1.03}$  & $   2.51_{ 2.10}^{ 3.10}$  \\
250\% Solar $\Omega_M=0.0$ $\Omega_{\Lambda}=0.0$  & 286.76 & 113.26 & $   1.08_{ 1.03}^{ 1.13}$  & $   4.36_{ 3.65}^{ 5.39}$  \\\\

~40\% Solar $\Omega_M=0.3$ $\Omega_{\Lambda}=0.7$  & 284.40 &  35.43 & $   1.07_{ 1.02}^{ 1.12}$  & $   1.36_{ 1.14}^{ 1.69}$  \\
100\% Solar $\Omega_M=0.3$ $\Omega_{\Lambda}=0.7$  & 311.23 &  80.37 & $   1.17_{ 1.12}^{ 1.23}$  & $   3.09_{ 2.59}^{ 3.83}$  \\
250\% Solar $\Omega_M=0.3$ $\Omega_{\Lambda}=0.7$  & 344.91 & 139.39 & $   1.30_{ 1.24}^{ 1.36}$  & $   5.36_{ 4.50}^{ 6.64}$  \\\\

\hline
\end{tabular}
\end{minipage}
\end{table*}

\vspace{1cm}

\appendix

\section{Calculation of the Luminosity-weighted Mean Metallicity of
a Spheroidal Galaxy Interior to an Isophotal Limit}

Elliptical galaxies are known to contain highly enriched cores with
strong metallicity gradients. Because of cosmological dimming, in the
present paper we sample ellipticals at a range of rest-frame limiting
isophotes, and it is important to have at least a qualitative
understanding of the effects of metallicity gradients on the sampled
starlight probed by our data. It is shown in Arimoto et al (1997) that
the mean luminosity-weighted iron abundance of spheroidal systems with
$R^{1/4}$ profiles integrated to infinity is similar to the abundance
measured at the effective radius. It is straightforward to generalize
the integral in equation (6) of Arimoto et al (1997) to calculate
$\langle Z(R)\rangle$, the mean metallicity of an $R^{1/4}$ law
spheroid within a given isophotal radius $R$, as a fraction of the
metallicity measured within the effective radius, $R_e$. For a
circularly symmetric galaxy the resulting expression is analytically
tractable.  The $R^{1/4}$ law giving surface brightness $I(R)$ as a
function of radius $R$ is parameterized as follows:
\begin{equation}
I(R) = \exp \left\{ -b \left[ {\left( R\over R_e \right)}^{1/4} -1 \right] \right\}
\end{equation}
where $R_e$ is the half-light radius, and $b=3.33\ln(10)$. Metallicity
gradients can be parameterized by a power-law index $c$ as a function
of radius:
\begin{equation}
Z(R) = Z(0)\left({R \over R_e}^{-c} \right).
\end{equation}
Essentially all of the ellipticals studied by Arimoto et al.  (1997)
have gradients parameterized by $0<c<1.3$, with metallicities of up to
several hundred percent solar in their cores decreasing to roughly
solar abundance at the effective radius.

Integrating equation (6) of Arimoto et al (1997) to a radius $R$
instead of to infinity yields the mean metallicity interior to a
radius R, expressed in units of the mean metallicity interior to
radius $R_e$:
\begin{equation}
\langle Z(R) \rangle  =  -e^{b (-1 + R^{1/4})} {A / B}\\
\end{equation}
where
\begin{eqnarray}
A & = & [-5040 b - 2520 b^2 - 840 b^3 - 210 b^4 - 42 b^5 \\
  &   & - 7 b^6 - b^7 + 5040 (-1 + E^b)] \cdot \nonumber \\
  &   & [\Gamma(8 - 4 c) - \Gamma(8 - 4 c, b R^{1/4})]  \\
B & = & [-5040 (-1 + e^{b R^{1/4}}) + 5040 b R^{1/4}\\
  &   & + 2520 b^2 \sqrt{R} + 840 b^3 R^{3/4} + 210 b^4 R + \nonumber \\
  &   & 42 b^5 R^{5/4} + 7 b^6 R^{3/2} + b^7 R^{7/4}] \cdot \\
  &   & [\Gamma(8 - 4 c) - \Gamma(8 - 4 c,b)].
\end{eqnarray}
(Note that in the preceeding expression, $\Gamma(x)$ is the Euler gamma function,
and $\Gamma(x,y)$ is the incomplete gamma function).

\begin{figure*}
\begin{center}
\leavevmode \centerline{\psfig{file=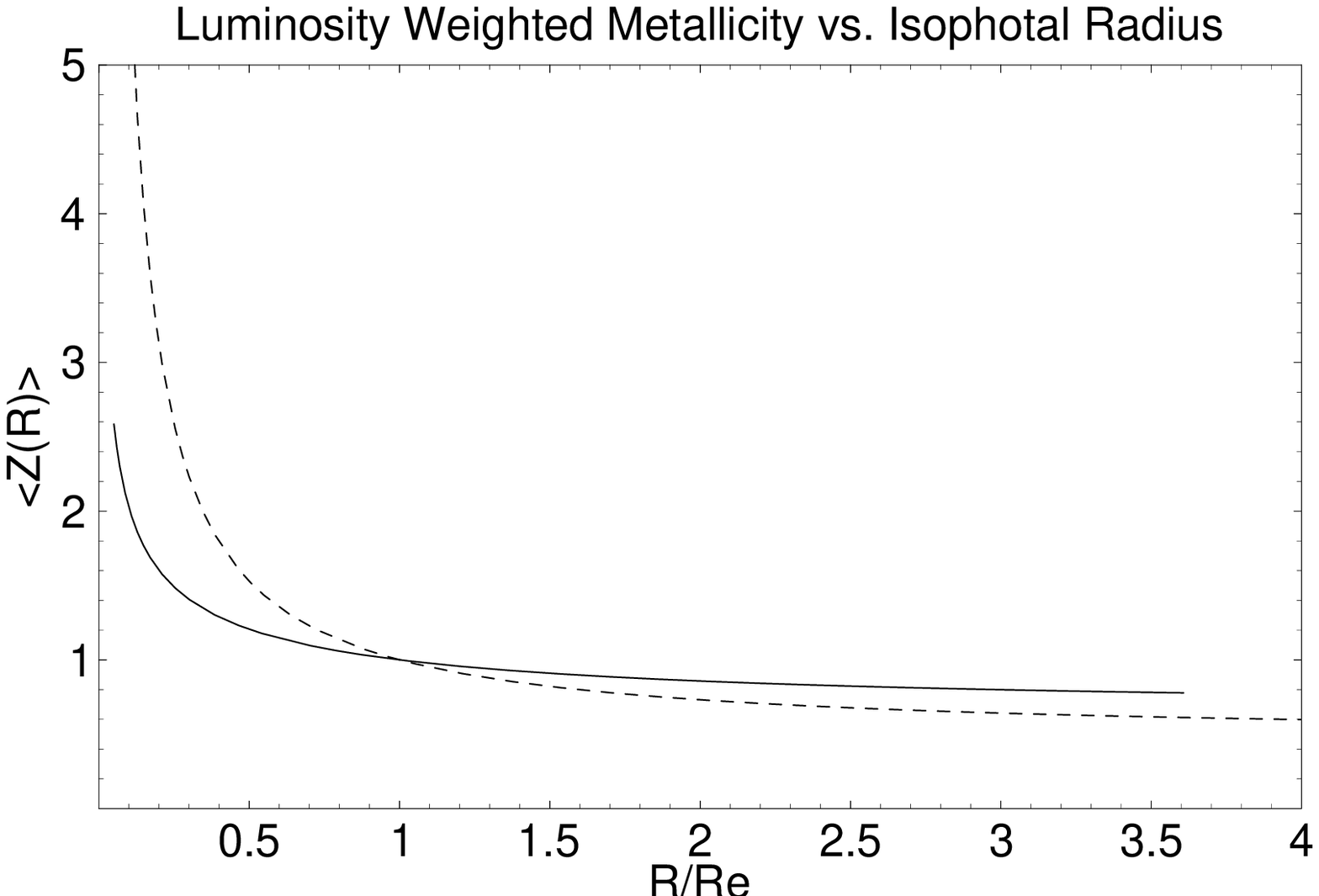,width=12cm,angle=0}}
\end{center}
\caption{\em The mean luminosity-weighted metallicity as a function of
isophotal radius. Metallicity is given as a fraction of the
metallicity interior to the half-light radius $R_e$. Radii are
expressed in units of the half-light radius. The dotted line
corresponds to $c=1.3$, and the solid line to $c=0.4$.}
\label{metal}
\end{figure*}

Mean metallicities as a function of radius are shown in
Figure~A1 for values of $c=0.4$ and $c=1.3$, spanning the
range of metallicity gradients typically seen in luminous
ellipticals. As expected, the luminosity-weighted metallicity as a
function of isophotal radius rises sharply at $R \ll R_e$, but for our
present purposes the important feature to note in Figure~12 is the
shallow decline beyond $R > 0.5 R_e$ in the luminosity-weighted mean
metallicity, regardless of the strength of the power-law slope $c$
parameterizing the metallicity gradients. This behaviour was also
noted by Arimoto et al. 1997 in their simplified calculation.  {\em
The overall metallicity of the galaxy can be well-characterized by the
mean metallicity interior to the effective radius (ie. $\sim$ solar in
the Arimoto et al. sample) over an extraordinarily large range in
isophotal radius.} We can thus make fair comparisons between
ellipticals over a broad range of redshifts in spite of cosmological
dimming of the rest-frame isophote, provided only the ellipticals are
observed with sufficient signal-to-noise that their isophotal radii
are at least as large as their effective radii. This is the case for
the vast majority of ellipticals in the present sample --- only at
very high redshifts, where compact ellipticals have limiting isophotes
comparable to their effective radii, is the mean luminosity-weighted
metallicity expected to deviate from that at the effective radius.

\label{lastpage}


\vspace{2cm}

\begin{thebibliography}{1}

\bibitem[]{} Abraham, R. G., Tanvir, N. R., Santiago, B. X., Ellis, R. S.,
Glazebrook, K. \& van  den Bergh, S. 1996a, MNRAS, 279, L47

\bibitem[Abraham et~al. 1996b]{Abraham-et.al-96}
Abraham, R.~G., van~den Bergh, S., Ellis, R.~S., Glazebrook, K.,
Santiago, B. X., Griffiths, R.~E., Surma, P. 1996b, ApjSS, 107, 1

\bibitem[]{} Abraham, R. G., Smecker-Hane, Tammy A., Hutchings, J. B.,
Carlberg, R. G., Yee, H. K. C., Ellingson, Erica, Morris, Simon, Oke,
J. B., Rigler, Michael, 1996c, ApJ, 471, 694

\bibitem[]{} Abraham,R. G., Ellis,R. S., Fabian, A. C., Tanvir, N. R. and
Glazebrook, K., 1998 MNRAS (astro-ph/9807140)

\bibitem[]{} Arimoto, N., Matsushita, K., Ishimaru, Y., Ohashi, T., \&
Renzini, A. 1997, ApJ, 477, 128

\bibitem[]{} Baade, W. 1957 in {\em Stellar Populations}, ed. O'Connell, D.J.K.
p3, Vatican (Rome).

\bibitem[]{} Barger, A. J., Aragon-Salamanca, A., Ellis, R. S., Couch, W. J.,
Smail, I., Sharples, R. M.,1996, MNRAS, 27, 1

\bibitem[]{} Barger, A. J., Cowie, L. L., Trentham, N., Fulton, E.,
Hu, E. M., Songalia, A. \& Hall, D., 1998, ApJ (in press)

\bibitem[]{} Baugh, C. Cole, S. \& Frenk, C.S. 1996 MNRAS 283, 1361.

\bibitem[]{} Bertin, E. \& Arnouts, S. 1996 Astron. Astrophys. Suppl. 117, 393.

\bibitem[]{} Bower, R.G., Lucey, J.R. \& Ellis, R.S. 1992 MNRAS 254, 601.

\bibitem[]{} Brinchmann, J., Abraham, R.G., Schade, D. et al 1998 Ap J 499, 112.

\bibitem[]{} Broadhurst, T.J., Ellis, R.S. \& Glazebrook, K. 1992 Nature

\bibitem[]{} Bruzual, G. \& Charlot, S. 1993 Ap J 405, 538

\bibitem[]{} Bruzual, G. \& Charlot, S. 1996, GISSEL (in preparation)

\bibitem[]{} Charlot, S. \& Silk, J., 1994, ApJ, 432, 453

\bibitem[]{} Connolly, A.J., Szalay, A.S., Dickinson, M. SubbaRao, M.U.
\& Brunner, R.J. 1997 Ap J 486, 11

\bibitem[]{} Cowie, A.S., 1997 unpublished catalogue of HDF redshifts
(see {\tt http://www.ifa.hawaii.edu/$\sim$cowie/tts/tts.html})

\bibitem[]{} Driver, S.P., Windhorst, R.A., Ostrander, E.J., Keel, W.C. et al
1995 Ap J 449, L23.

\bibitem[]{} Dunlop, J. 1998 in {\em The most distant radio galaxies}, eds.
Best, et al, Kluwer, in press (astro-ph/9801114)

\bibitem[]{} Efron, B. \& Tibshirani, R. J. 1993. ``An Introduction to the
Bootstrap'', (Chapman \& Hall:New York)

\bibitem[]{} Ellis, R.S. Smail, I., Dressler, A. et al 1997 Ap J 483, 582

\bibitem[]{} Glazebrook, K., Ellis, R.S., Santiago, B. \& Griffiths, R.E. 1995
MNRAS 275, L19.

\bibitem[]{} Glazebrook, K., Abraham, R., Santiago, B., Ellis, R.S., \&
Griffiths, R. 1998 MNRAS 297, 885.

\bibitem[]{} Governato, F., Baugh, C.M., Frenk, C.S. et al 1998 Nature 392, 359

\bibitem[]{} Groth, E., Kristian, J.A., Lynds, R. et al 1994 Bull. AAS 185, 53.09

\bibitem[]{} Hogg, D.W., Cohen, J.G., Blandford, R. et al 1998 AJ 115, 1418

\bibitem[]{} Im, M \& Casertano, S. 1998 preprint

\bibitem[]{} Im, M. Griffiths, R.E., Ratnatunga, K.U. \& Sarajedini, V.L.
1996 Ap J 461, L9

\bibitem[]{} Jimenez, R., Friaca, A., Dunlop, J., Terlevich, R.,
Peacok, J. \& Nolan, L. 1999 MN submitted (astro-ph 9812222)

\bibitem[]{} Kauffmann, G., Charlot, S. \& White, S.D.M. 1996 MNRAS 283, L117.

\bibitem[]{} Kauffmann, G. \& Charlot, S. 1998a MNRAS 297, L23.

\bibitem[]{} Kauffmann, G. \& Charlot, S. 1998b, preprint (astro-ph/9810031).

\bibitem[]{} Lilly, S.J., Tresse, L., Hammer, F., Crampton, D. \& LeF\'evre, O.
1995 Ap J 455, 108

\bibitem[]{} Lilly, S.J., LeF\'evre, O., Hammer, F. \& Crampton, D. 1996
Ap J 460, L1.

\bibitem[]{} Marzke, R.O., da Costa, L.N., Pellegrini, P.S., Willmer, C.N.A.
\& Geller, M.J. 1998 Ap J in press (astro-ph/9805218)

\bibitem[]{} Menanteau, F., Abraham, R.G. \& Ellis, R.S. 1999 (in preparation)

\bibitem[]{} Moustakas, L.A., Davis, M., Graham, J. \& Silk, J. 1997, ApJ,
475, 445.

\bibitem[]{} Peacock, J.A., Jimenez, R., Dunlop, J., Waddington, I. et al
1998 MNRAS in press (astro-ph/9801184)

\bibitem[]{} Perlmutter, S. et al 1999 ApJ, in press
(astro-ph/9812133)

\bibitem[]{} Pozzetti, L. Bruzual, G.,  Zamorani, G., 1996 MNRAS 281, 953

\bibitem[]{} Rose, J., Bower, R.G., Caldwell, N. et al 1994 AJ 108, 2054

\bibitem[]{} Sandage, A. 1986 in {\it Deep Universe}, ed. Binggeli, B. \&
Buser, R. p1, Springer-Verlag (Berlin).

\bibitem[]{} Sandage, A., Visvanathan, N. 1978 Ap J 223, 707

\bibitem[]{} Schade, D. Lilly, S.J., Crampton, D. et al 1998 in preparation

\bibitem[]{} Stanford, S.A., Eisenhardt, P.R. \& Dickinson, M. 1998 Ap J 492, 461.

\bibitem[]{} Totani, T. \& Yoshii, Y. 1998 preprint (astro-ph/9805262)

\bibitem[]{} van den Bergh, S., Abraham, R. G., Ellis, R. S., Tanvir, N. R.,
 Santiago, B. X., Glazebrook, K. G.,1996, AJ, 112, 359

\bibitem[]{} van Dokkum, P., Franx, M., Kelson, D.D. et al 1998 Ap J, in press
(astro-ph/9807242)

\bibitem[]{} Wang, Y., Bahcall, N. \& Turner, E.L. 1998 AJ in press (astro-ph/
9804195).

\bibitem[]{} Williams, R.E., Blacker, B., Dickinson, M. et al 1996 AJ 112, 1335.

\bibitem[]{} Windhorst, R., Driver, S.P., Ostrander, E.J. et al 1996 in
{\it Galaxies in the Young Universe}, ed. Hipperlein, H., p265,
Springer-Verlag (Berlin).

\bibitem[]{} Zepf, S.E. 1997, Nature, 390, 377
\end{thebibliography}
\end{document}